# Fe ion doping effect on electrical and magnetic properties of $La_{0.7}Ca_{0.3}Mn_{1-x}Fe_xO_3$ ($0 \leq x \leq 1$)

Neeraj Kumar, H. Kishan and V.P.S. Awana*

*Superconductivity and Cryogenics Division, National Physical Laboratory*
*Dr. K.S. Krishnan Marg, New Delhi-110012*

## Abstract

Here we report the structural, electrical and magnetic properties of Fe doped $La_{0.7}Ca_{0.3}Mn_{1-x}Fe_xO_3$ (LCMFO; $0 \leq x \leq 1$) prepared by conventional solid state reaction method. Simulated data on XRD shows an increase in volume with an increase in Fe ion concentration. XPS supports that $Fe^{3+}$ ions directly substitute $Mn^{3+}$ ions. Shifting towards lower wave-number and symmetric IR band structure confirms increase in volume and homogeneous distribution of Fe ions. Fe ion doesn't contribute in double-exchange (DE) conduction mechanism due to its stable half filled 3d orbital. The presence of $Fe^{3+}$ ions encourages anti-ferromagnetism (AFM) generated by super-exchange interaction and suppress insulator-metal transition temperature ($T_{IM}$). Magnetic measurements show the existence of magnetic polarons supported by increase in volume of unit cell and deviation from Curie-Weiss law.

PACS: 75.47.Lx, 71.38.Ht, 71.30.+h

Key words: Manganites, Metal–insulator transition, Super exchange interaction.

Corresponding author's address: Dr. V.P.S. Awana
National Physical Laboratory, Dr. K.S. Krishnan Marg, New Delhi-110012, India
Fax No. 0091-11-45609310: Phone no. 0091-11-45609210
e-mail-awana@mail.nplindia.ernet.in: www.freewebs.com/vpsawana/

# 1. INTRODUCTION

Transition-metal doped oxides have been the subject of study since 1950`s because of their unique electrical, magnetic and structural properties [1]. But in the recent years a new class of these oxides termed as manganite perovskite attracts more attention [2]. In 1993 von Helmolt et al noticed a huge change in electrical resistivity on the application of magnetic field and this phenomenon was named as colossal magneto-resistance (CMR) and may play important role in several areas like magnetic sensors, magnetic data storage devices [3]. However, magneto-resistance is also found in other class of transition-metal oxides as pyrochlores and spinels [4]. However, in manganite materials lattice and electron interactions are very strong and generate more complex physical phenomena by the variation of chemical composition, temperature and magnetic field. Such manganite materials have general formula $ABO_3$, where A is a trivalent rare-earth ion like $La^{3+}$, $Pr^{3+}$, $Nd^{3+}$ or divalent alkali-earth ion like $Ca^{2+}$, $Sr^{2+}$, $Ba^{2+}$ etc. and B is the manganese ion, which can also be replaced by other transition metal ion. If A is replaced by divalent alkali-earth ion, there exist $Mn^{3+}/Mn^{4+}$ pairs in the proportions of rare-earth ion and alkali-earth ion. Due to the presence of $Mn^{3+}/Mn^{4+}$ ions, at a particular range of doping an insulator-metal transition ($T_{IM}$) accompanied by a paramagnetic- ferromagnetic transition can be observed. When B-site is replaced by a transition metal ion having nearly similar ionic size one can suppress the change in physical properties arising due to distortion of $MnO_6$ octahedra. The transport properties of the manganite materials are explained by the double-exchange and electron-phonon coupling arising from Jahn-Teller distortion. This Jahn-Teller (JT) distortion generates lattice polarons that localizes the carriers and hence results an increase in the resistivity [5]. Moreover, these lattice polarons may be either small-range or variable range polarons. However, experimental results support that manganite are magnetically inhomogeneous. So that concept of lattice polarons may no longer can go alone and magnetic polarons was proposed on the basis of spin polarized neutron scattering and Mossbauer spectroscopy [6]. In magnetic polarons carrier is trapped along with spin alignment in a particular direction by an ionic site. Above $T_{IM}$ these magnetic polarons are precipitated in paramagnetic matrix. By the application of temperature or magnetic field these magnetic polarons can grow and generates ferromagnetic phase. Whenever, the manganite have small lattice distortion, then it is very difficult to identify which polarons i.e. lattice or magnetic are responsible for the change in physical properties.

Therefore, it is of utmost importance that a direct observation of magnetic polarons must be seen in a minimal distorted system like with Fe, Cr-doping. Fe, Cr atom is nearest neighbor to Mn atom in the periodic table having nearly same radius results so a little distortion is expected. Also Mn, Fe and Cr have similar energy level structure and multiple valences. As $Cr^{3+}$ ion has similar electronic structure as $Mn^{4+}$, there exists a possibility of double-exchange between them reported by Maignan et al. [7] for $Sm_{0.56}Sr_{0.44}Mn_{1-x}Cr_xO_3$ system and Raveau et al. [8] for $Pr_{0.5}Ca_{0.5}MnO_3$. On the other hand $Fe^{3+}$ ion has complete 3d filled orbital, so it does not play a part in double-exchange. Ahn et al. [9] observed that Fe doping at Mn site encourages the anti-ferromagnetic behavior and suppression of the double exchange effects. However ferromagnetism due to Fe doping is less pronounced. Also Mn-O-Mn and Fe-O-Fe chains coexist that encourages magnetic inhomogeneity. Cai et.al observed ferromagnetic clusters embedded in anti-ferromagnetic clusters for LCMO doped with Fe (10%) and found suppression in double exchange [10]. J M Liu et. al reported the transport properties of Fe doped $Pr_{0.75}Sr_{0.25}MnO_3$ (PSMO) and found obeying variable range hopping conduction mechanism. Above transition temperature magnetic polarons are percolated in insulating paramagnetic phase [11].

As there are few reports for the whole Fe doping concentration, in this work, we try to investigate the effect of Fe doping on structural, transport and magnetic properties of $La_{0.7}Ca_{0.3}Mn_{1-x}Fe_xO_3$ ($0 \leq x \leq 1$) compound. Fe ions and Mn ions both are magnetic, also $Fe^{3+}$ can exist in high and low spin states. It may be interesting to study, if Fe can exist in $Fe^{4+}$ state or due to fluctuating spin state, it can take part in double-exchange.

## 2. EXPERIMENTAL

The samples in the series with nominal composition $La_{0.7}Ca_{0.3}Mn_{1-x}Fe_xO_3$ ($0 \leq x \leq 1$) were prepared by conventional solid state reaction method. Stoichiometric ratio of $La_2O_3$, $CaCO_3$, $Fe_2O_3$ and $MnO_2$ (all from Aldrich chemical Ltd with 99.9% purity) were mixed very well to obtain a homogeneous mixture. The mixtures were calcined at 1250 $^0C$ for 24h. The calcined mixtures were then pressed into pellets with the addition of small sum of poly vinyl acetate (PVA) and sintered in air at 1325 $^0C$ over 36h. Finally the pellets were annealed for 12 hours at 1100 $^0C$ in oxygen atmosphere for the removal oxygen in-homogeneity and strain induced during sintering. The structural information was obtained by XRD studies on RIGAKU machine with Cu-Kα radiation (1.54 Å) in the 2θ range of $20^0$-$80^0$. XPS analysis was performed with Mg-Kα source (1253.6 eV). Spectra were taken after purging the sample at a vacuum pressure of ~$10^{-10}$ torr. Mn 2p core level binding energies were calculated with regard to Carbon 1s (C1s) peak fixed at 284.8 eV. Electrical resistivity was measured as a function of temperature by conventional four probe method and IR spectra were recorded on Nicolet FTIR 5700 spectrometer.

Magnetic measurements for $La_{0.7}Ca_{0.3}Mn_{1-x}Fe_xO_3$ compound were carried out using SQUID magnetometer (Quantum Design, MPMS).

## 3. RESULTS AND DISCUSSIONS

A first direct evidence of phase formation and Fe substitution can be seen from X-ray diffraction spectra as shown in Figure 1, where no extra reflection is observed in any spectra. Similar to parent compound all substituted perovskite are monophasic, implying that Mn has been completely substituted by Fe. All the samples with composition $La_{0.7}Ca_{0.3}Mn_{1-x}Fe_xO_3$ ($0 \leq x \leq 1$) are indexed single phase orthorhombic structure with *Pbnm* space group. The pristine compound LCMO has been indexed with lattice parameters as 5.462(4) Å, 5.478(4) Å and 7.721(4) Å. However, a regular shift towards lower angle is noticed on increasing Fe content. On the contrary, many research groups have reported no appreciable shift with Fe-doping [9, 12]. Shift in peaks towards lower angles indicates an increase in the volume which is confirmed by Rietveld refinement data also as shown in Table 1. Although, volume is a complex quantity as in $La_{1-x}Sr_xMnO_3$ there is a decrease in volume except Sr ion has larger ionic radii [13]. As $Mn^{3+}$ ion and $Fe^{3+}$ ion have same ionic radii (0.0645 *nm*), so that one possibility of increase in volume may be due to the presence of $Fe^{4+}$ ions because $Fe^{4+}$ ions have larger ionic radii (0.0585 *nm*) than $Mn^{4+}$ ion (0.053 *nm*) [14].

For the possible presence of $Fe^{4+}$ ions, we carried out an X-ray-photoemission spectra for $La_{0.7}Ca_{0.3}Mn_{1-x}Fe_xO_3$ (x=0.3, 0.5) are shown in Figure 2. The binding energies and peaks match well with the earlier reported data [15]. The inset of figure 2 shows the Mn 2p core level spectra of the respective samples with the expected multiplet splitting. For the analysis of Mn 2p core level spectra a linear background was used. These experimental spectra were fitted with two peaks along with a large peak representing satellite. The peak maxima corresponds with $2p_{3/2}$ and $2p_{1/2}$ are at (641.9 *eV*, 653.9 *eV*) and (642.1 *eV*, 653.9 *eV*) for x = 0.3, 0.5 respectively. The $2p_{3/2}$ peak corresponding to $MnO_2$ is observed at 642.6 *eV* with Mn ion in 4+ valence state, on the other hand, for $Mn_2O_3$ 3+ valence state this peak is observed at 641.6 *eV*. In the present case, corresponding peaks are observed in between these two values and very well consistent with reported earlier by Taguchi and Shimida [16]. This supports that in our case Mn is present in both 3+ and 4+ ionic state. Gaussian fitted component indicate the presence of both $Mn^{3+}$ and $Mn^{4+}$ ions. As $Fe^{3+}$ ions directly replaces $Mn^{3+}$ ions, the calculated ratio of $Mn^{3+}/Mn^{4+}$ is 0.91 and 0.68 for x = 0.3, 0.5, doped compound, which is close to the expected ratio 1.33 and 0.67 respectively.

In conclusion as we increase Fe concentration $2p_{3/2}$ peak shifts towards higher energy which is attributed due to the presence of $Mn^{4+}$ ions along with reduction in $Mn^{3+}$ ions. Fe doping directly replaces $Mn^{3+}$ ion with $Fe^{3+}$ valence state as reported earlier studies by K. Ghosh et.al [9, 17]. Hence relatively $Mn^{4+}$ ions percentage remains nearly invariant.

Ignoring the presence of $Fe^{4+}$ ions, other possibility may be the formation of magnetic polaron (self trapping of carrier along with lattice distortion) that can affect Mn-O bond length results a change in volume. S. Fillho et al. [18] have reported the presence of magnetic polaronic state in Fe doped LSMO system as a consequence of shift in XRD peaks towards lower angles. Very similar behavior are found in XRD pattern for our samples also, we will check out the possible reasons of the increase in lattice volume with Fe-doping in connection with electrical and magnetic properties.

Figure 3 shows normalized resistivity variation of temperature of the nominal series compounds $La_{0.7}Ca_{0.3}Mn_{1-x}Fe_xO_3$. The parent compound $La_{0.7}Ca_{0.3}MnO_3$ shows an metal– insulator transition ($T_{MI}$) at 257$K$ corresponding very well with earlier published data [1, 4, 5]. As we increase the Fe doping concentration in the parent compound, $T_{MI}$ temperature shifts towards low temperatures with an overall increase in the electrical resistivity. With Fe doping the resistivity of the parent compound (0.27 $\Omega$-cm) increases by almost four hundred times (130.7 $\Omega$-cm) of magnitude only with small amount of doping of x = 0.08. The remarkable influence of Fe doping on electrical transport properties is characterized by the tremendous shift of $T_{MI}$ temperature from 257$K$ to 90.4$K$ only for x = 0.08 and destruction of conductivity with Fe doping. Such large sensitivity to Fe doping suggests that Fe-doping suppresses the double-exchange conduction mechanism in a tremendous way [19]. As XPS study suggests the presence of only $Fe^{3+}$ ions in our system and due to their half filled stable 3d-state ($3d^5$, $t^3_{2g}$ $e^2_g$) Fe doesn't takes part in double-exchange (DE) conduction mechanism. With Fe doping the available hopping sites suppresses and x > 0.08 compound behaves as insulator as shown in figure 3.

The inset of figure 3 shows the fitting of resistivity data for x < 0.1 with temperature by Holstein's model (small range hopping conduction) [20] in adiabatic regime above $T \sim \theta_D/2$:

$$\rho = \rho_0\, T\, exp\, (E_\rho / k_B T) \qquad (1)$$

Where $\rho_0$ is a constant, $E_\rho$ is the polaron hopping energy, and $k_B$ is the Boltzmann constant. Above the Debye temperature ($\theta_D$) the conduction mechanism is accounted for thermally activated polaron hopping. The calculated values of hopping energy and Debye temperature ($\theta_D$) are listed in Table 2. This is surprising that with Fe doping hopping energy decreases in contrast to the increase in the electrical resistivity. These results can be viewed in terms of the JT distortion of the $Mn^{3+}$ ions. In pristine material nearly 70% $Mn^{3+}$ ions are present which JT active are. Booth et al. plotted a linear relationship between Mn-O distortion and $Mn^{3+}$ ion concentration and J. M. De Teresa et al. find a linear behavior of polaron binding energy with $Mn^{3+}$ ion content [21, 22]. From these findings we can conclude that with an increase in $Mn^{3+}$ ion concentrations, charge carriers get localized and result in the corresponding increase in hopping energy. As $Fe^{3+}$ ions directly replace $Mn^{3+}$ ions, a reduction in JT distortion is expected due to $Fe^{3+}$ being JT inactive ion. Also, with it's no participation in the DE interaction the resistivity increases by several orders. In parent compound (LCMO) $Mn^{+3}$-$O^{-2}$-$Mn^{+4}$ networks is distributed homogeneously and conduction follow double-exchange mechanism. With Fe doping this network breaks and Fe is situated in either $Fe^{3+}$ – $O^{2-}$ - $Fe^{3+}$ or $Mn^{3+}$ -$O^{2-}$ - $Fe^{3+}$ networks. In both cases double-exchange network gets weakened and results an increase in resistivity.

Figure 4 shows the temperature coefficient of resistance (TCR calculated as *(1/R) (dR/dT))* values for samples $La_{0.7}Ca_{0.3}Mn_{1-x}Fe_xO_3$ ($0 \leq x \leq 0.08$). For the parent compound TCR is ~15% and remains same with increasing Fe concentration. TCR represents the sharpness of $T_{MI}$ transition and believed to be dependent on inter-grain tunnel resistance. Large value of TCR can be obtained by producing conducing grain boundary by metal that have low melting temperature, so that it connects two grains at grain boundary. Nearly same value of TCR could be attributed to ensuing morphological effects in terms of grain size etc, instead of pure electronic process.

Inset of Figure 4 shows the IR spectra of parent (LCMO) and end member (LCFO) of the series in the mid IR range of 415 $cm^{-1}$ to 650 $cm^{-1}$ at room temperature. The present samples crystallize in orthorhombic structure as a distortion of the $MnO_6$ octahedral from their position in ideal cubic structure. Out of 20 vibration modes, only eight are IR active modes [23]. There is a shift in the vibration mode from 586.4 $cm^{-1}$ for the LCMO to 572.55 $cm^{-1}$ for LCFO compound. Shift in vibration mode towards lower wave number side indicates an increase in volume of unit cell and corroborates the Rietveld refinement analysis.

Figure 5 shows the magnetization variation with temperature of the series compounds. The ferromagnetic to paramagnetic transition temperature ($T_C$) calculated by *dM/dT* is nearer to transition temperature found in Resistivity variation measurements. With Fe - doping $T_C$ decreases in similar fashion as $T_{MI}$ in the electrical resistivity measurement. As we increase Fe-doping concentration the magnetization moment decreases due to increases in anti-ferromagnetic interaction. From magnetization variation with temperature data it can be clearly seen that for x = 0.40, there is anti-ferromagnetic to paramagnetic transition. The experimental observed transition temperatures for nominal series are listed in table 2. To confirm this assumption, we fitted our data using curie-Weiss law

$$\chi = C / (T - \theta_C) \qquad (2)$$

where $\chi$ is magnetic susceptibility, $\theta_C$ is Curie temperature coefficient and C is the constant that is related to the effective magnetic moment. Figure 6 shows the fitted data using Eqn. 2. Curie law fits well for the parent compound up to transition temperature. But with Fe doping, the observed data depart from this equation. This deviation is usually thought of cluster-spin glass state, in which no long range order is present. More-over, ferromagnetic clusters are embedded within anti-ferromagnetic clusters and conduction takes place through percolation mechanism [24]. Also for more accuracy we carried out magnetization (M) under zero field and field cooled conditions for x = 0.40 and 1.0 concentration. With an increase in spin glass behavior ZFC and FC magnetization separation should be increase. In figure 7, this behavior can be seen clearly along with anti-ferromagnetic-paramagnetic transition. Magnetization studies have described LCFO as a canted-anti-ferromagnet with ordering temperature of above 400 $K$.

In order to further probe the magnetic properties magnetization versus field (*M-H*) measurements is carried out at 5$K$ as shown in inset of Figure 8. We tried to make a rough estimation for the prediction of magnetic moment as $Mn^{3+}$ ions is replaced by $Fe^{3+}$ ions. The chemical compound with the chemical formula $La^{3+}_{0.7}Ca^{2+}_{0.3}(Mn_{0.7-x}Fe_x)^{3+}(Mn)^{4+}_{0.3}O_3$ leads to a magnetic moments of

$$M(\mu_s) = [4(0.7 - 0.7x) - 5(0.7x) + 3(0.3)] \, \mu_B$$

$$= (3.7 - 6.3x) \, \mu_B \qquad (3)$$

As $Mn^{3+}$, $Mn^{4+}$ and $Fe^{3+}$ ions have magnetic moment ~ 4$\mu_B$, 3$\mu_B$ and 5$\mu_B$ respectively. The calculated and experimental values of magnetic moment per formula unit are listed in Table 2. There is small discrepancy in them and difference increase with Fe concentration as spin glass behavior increases. The *M* for the parent compound saturates with magnetic moment ~3.96 $\mu_B$. The observed value of magnetic moment is higher than the calculated value (3.7 $\mu_B$). It may be due to some oxygen homogeneity or may be attributed by the La-Mn interaction [25]. Clearly the occurrence of hysteresis loop along with linear nature and small magnetic moment implies the presence of FM interaction with AFM interaction that dominates with Fe doping. For x>0.30 magnetic measurements shows an anti-ferromagnetic nature with Neel temperature below 50 *K*. For all the other doped compounds *M* does not show saturation even under the field of 20 *KOe*. The qualitative magnetic behavior of $La_{0.7}Ca_{0.3}Mn_{1-x}Fe_xO_3$ system is sum up in main panel of Figure 8. The $La_{0.7}Ca_{0.3}Mn_{1-x}Fe_xO_3$ transforms from pure ferromagnetic (FM) for lower x (x<0.10) values to AFM for higher x (x>0.30) and exhibits spin glass (SG) like character for intermediate x i.e. $0.10 < x < 0.30$. The SG reason appears mainly due to competing DE driven FM and SE driven AFM interactions in the studied $La_{0.7}Ca_{0.3}Mn_{1-x}Fe_xO_3$ system.

## 4. CONCLUSION

Structural, electrical and magnetic properties of Fe doped $La_{0.7}Ca_{0.3}Mn_{1-x}Fe_xO_3$ ($0 \leq x \leq 1$). XRD reveals that Fe ions substitute Mn ion completely. Using Reitveld refinement an increase in the lattice volume is observed with Fe-doping attributed to the formation of magnetic polarons. From magnetic measurements it is suggested that Fe ion doesn't take part in double-exchange mechanism, on the other hand it encourages super-exchange generated anti-ferromagnetism.

**Acknowledgement**

Neeraj Kumar is thankful to Council of Scientific and Industrial Research (CSIR), India and further acknowledges the financial support from CSIR in terms of CSIR-SRF Fellowship to carry out research work at NPL and pursue for his Ph.D thesis. Authors also acknowledge Dr. Amish joshi for XPS measurements.

**Figure/Table Captions**

Figure 1. XRD pattern for the series $La_{0.7}Ca_{0.3}Mn_{1-x}Fe_xO_3$ ($0 \leq x \leq 1$) compounds and Reitveld fitted pattern of $La_{0.7}Ca_{0.3}MnO_3$ and $La_{0.7}Ca_{0.3}FeO_3$

Figure 2. Overall XPS spectra of $La_{0.7}Ca_{0.3}Mn_{1-x}Fe_xO_3$ (x = 0.03, 0.05) (Inset set Shows the fitted Mn 2p core level spectra for x = 0.03 and 0.05 samples)

Figure 3. Temperature ($K$) variation of resistivity (Normalized) for the series $La_{0.7}Ca_{0.3}Mn_{1-x}Fe_xO_3$ ($0 \leq x \leq 0.08$)

Figure 4. Temperature coefficient of resistance variation for the series $La_{0.7}Ca_{0.3}Mn_{1-x}Fe_xO_3$ ($0 \leq x \leq 0.08$). (Inset shows the IR spectra for the series $La_{0.7}Ca_{0.3}Mn_{1-x}Fe_xO_3$ (x = 0.0, 1.0)

Figure 5. Temperature ($K$) variation of Magnetization for the series $La_{0.7}Ca_{0.3}Mn_{1-x}Fe_xO_3$ ($0 \leq x \leq 1$)

Figure 6. Curie-Weiss law fitted inverse susceptibility variation for the series $La_{0.7}Ca_{0.3}Mn_{1-x}Fe_xO_3$ ($0 \leq x \leq 0.3$)

Figure 7. Temperature ($K$) variation of Magnetization under ZFC and FC conditions for the series compound $La_{0.7}Ca_{0.3}Mn_{0.6}Fe_{0.4}O_3$ and $La_{0.7}Ca_{0.3}FeO_3$

Figure 8. Magnetic phase diagram of the series $La_{0.7}Ca_{0.3}Mn_{1-x}Fe_xO_3$ ($0 \leq x \leq 1.0$) Compound (In set shows the magnetization variation with applied external field for the series $La_{0.7}Ca_{0.3}Mn_{1-x}Fe_xO_3$ ($0 \leq x \leq 0.3$)

Table 1. Unit-cell parameters and Reliability Factor ($\chi^2$) for the refinements of $La_{0.7}Ca_{0.3}Mn_{1-x}Fe_xO_3$ ($0 \leq x \leq 1$) phases in *Pbnm* Space Group

Table 2. Fitting parameters of resistivity vs. temperature curves charactering small range polaron fits in high temperature regime

**Figure 1**

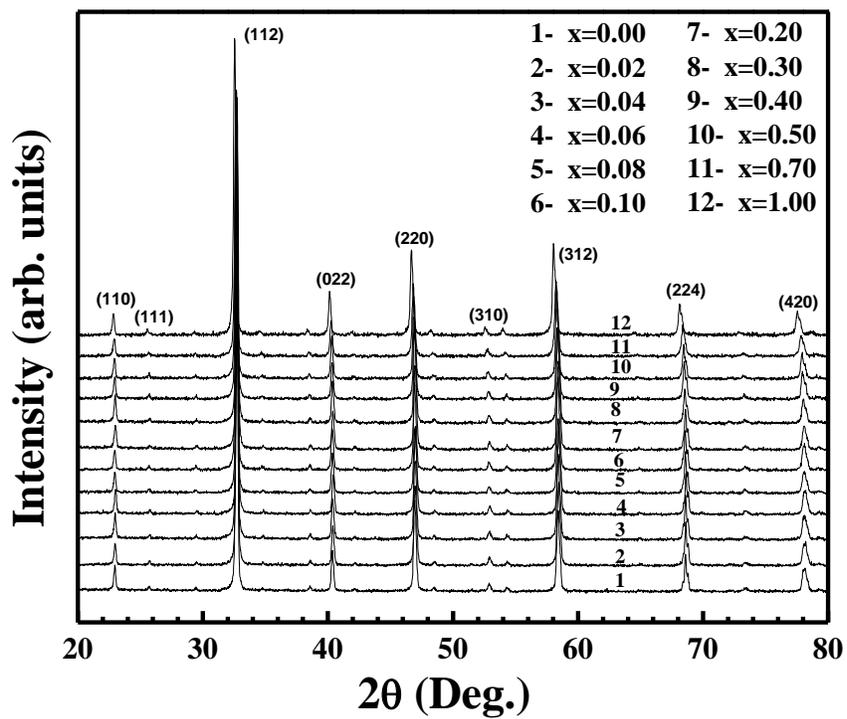

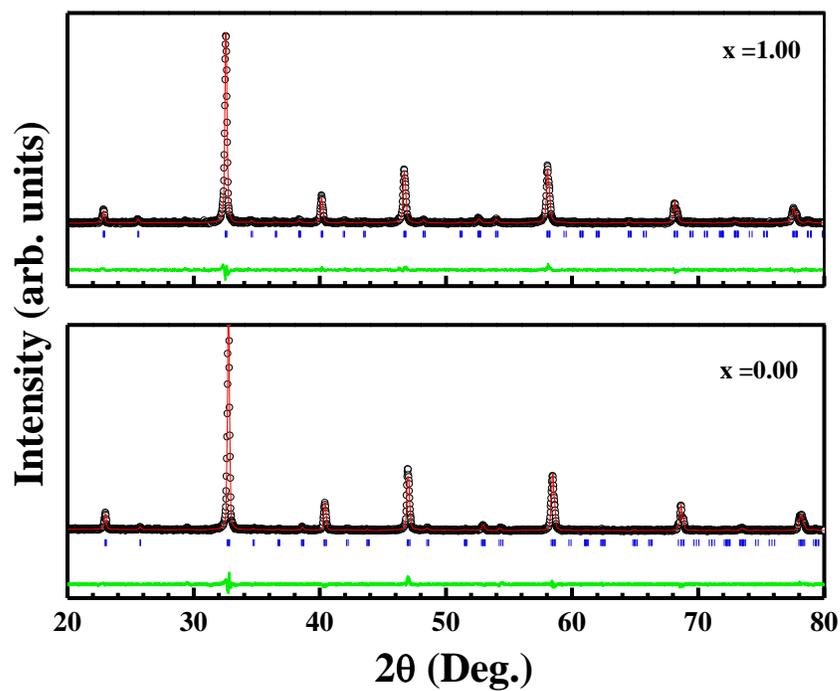

**Figure 2**

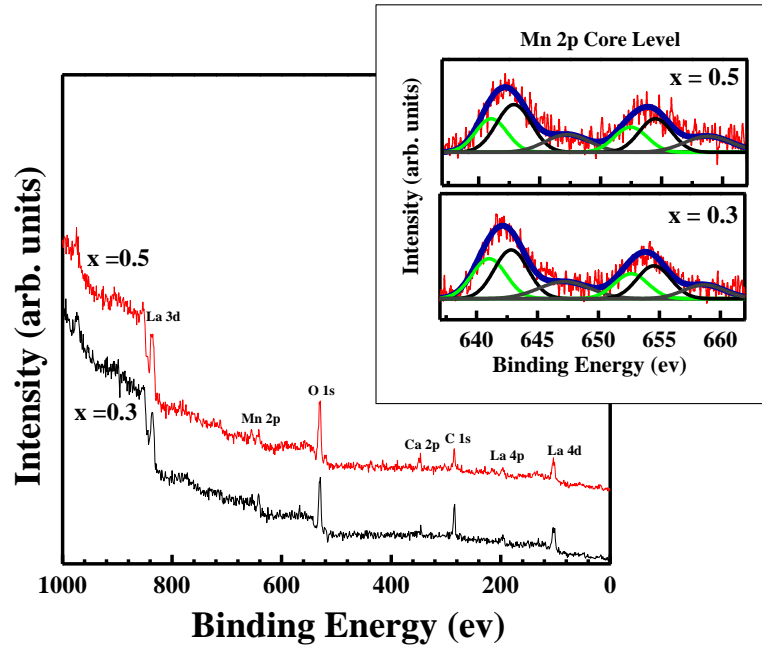

**Figure 3**

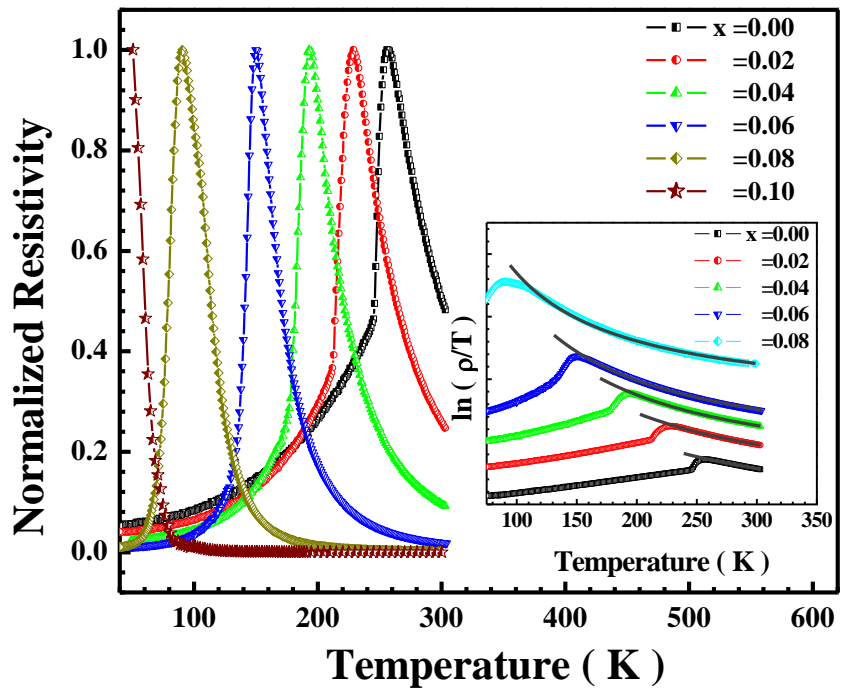

**Figure 4**

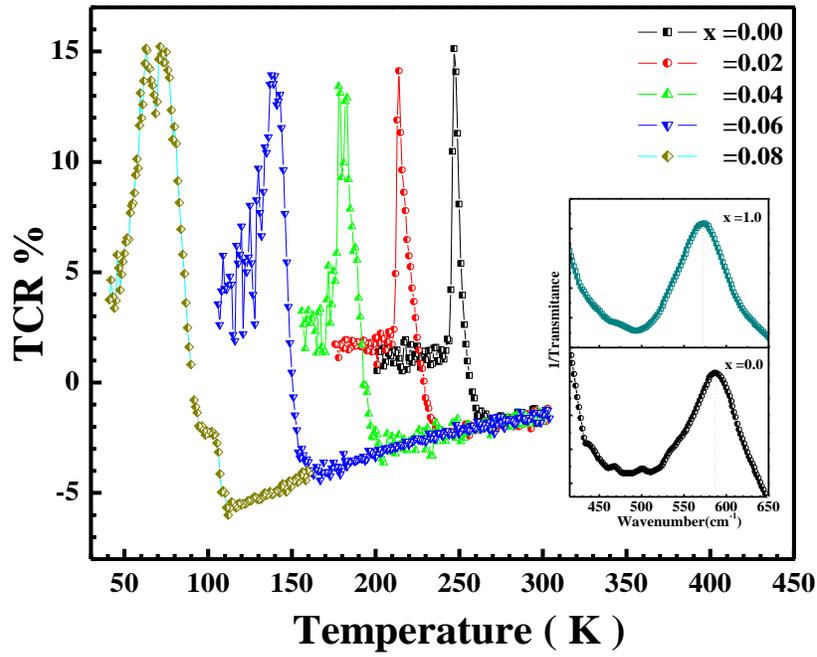

**Figure 5**

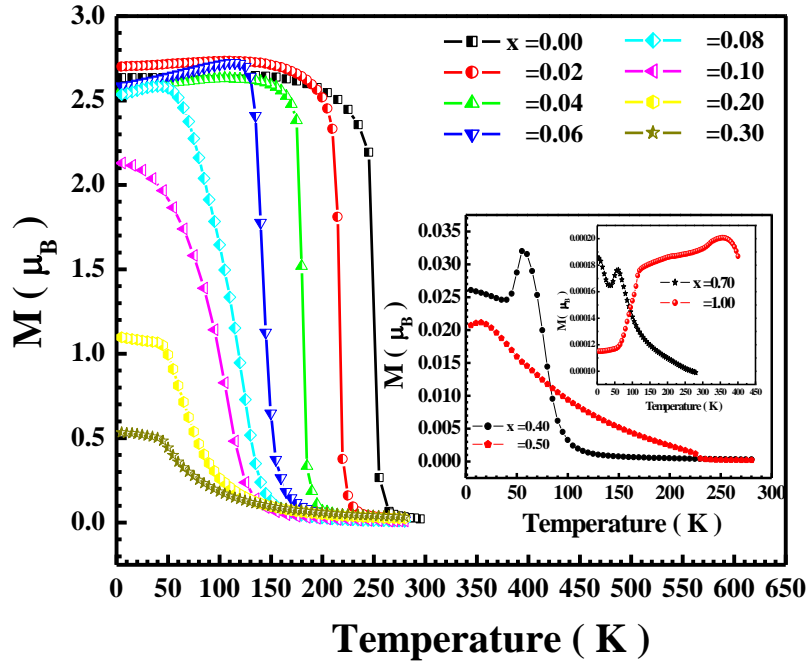

**Figure 6**

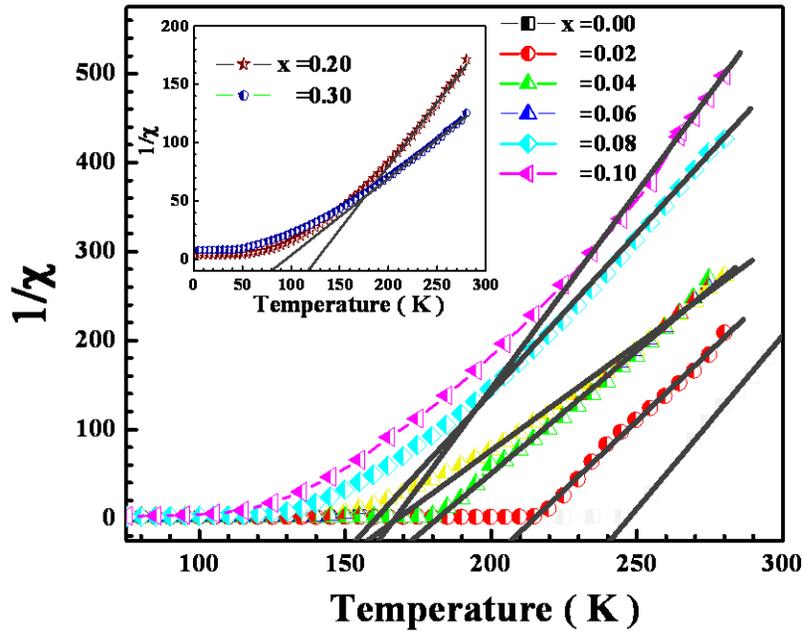

**Figure 7**

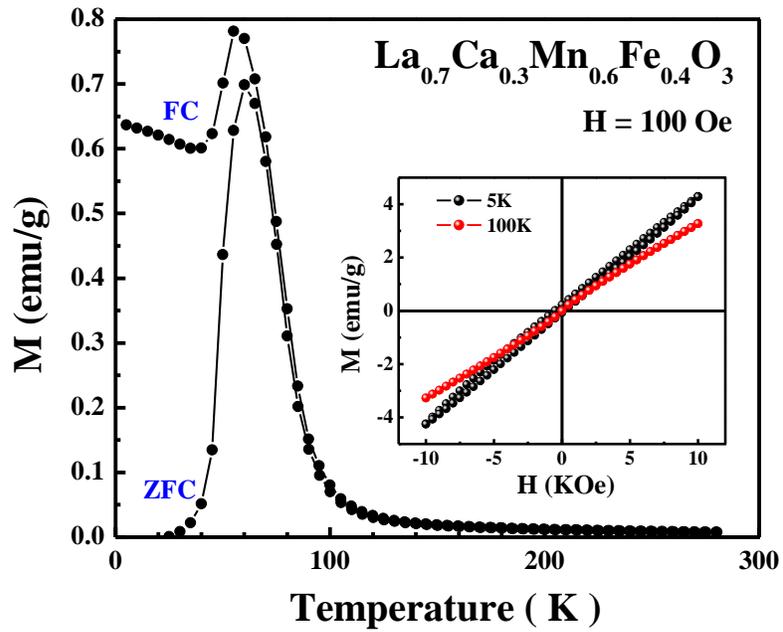

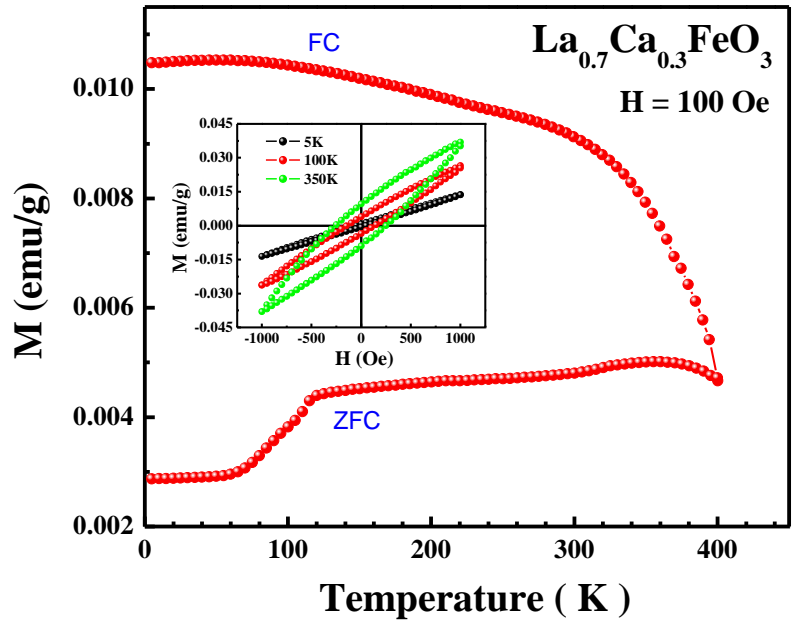

**Figure 8**

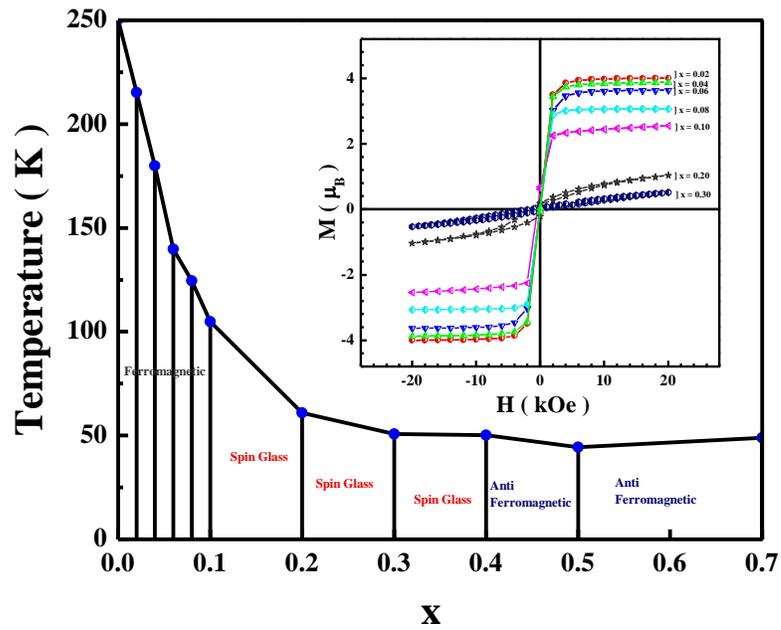

**Table 1**

| Sample ID ($L_{0.7}Ca_{0.3}Mn_{1-x}Fe_x O_3$) | a (Å) | b (Å) | c (Å) | $\chi^2$ | Volume (Å$^3$) |
|---|---|---|---|---|---|
| x =0.00 | 5.462(4) | 5.478(4) | 7.721(4) | 2.46 | 231.04(3) |
| x =0.02 | 5.463(4) | 5.478(4) | 7.721(4) | 2.26 | 231.13(3) |
| x =0.04 | 5.463(4) | 5.478(4) | 7.723(4) | 2.31 | 231.19(3) |
| x =0.06 | 5.464(4) | 5.478(4) | 7.724(4) | 2.35 | 231.26(3) |
| x =0.08 | 5.464(4) | 5.478(4) | 7.723(4) | 2.25 | 231.22(3) |
| x =0.10 | 5.464(4) | 5.477(4) | 7.723(4) | 2.35 | 231.20(3) |
| x =0.20 | 5.467(4) | 5.480(4) | 7.729(4) | 2.26 | 231.63(3) |
| x =0.30 | 5.470(4) | 5.481(4) | 7.734(4) | 2.35 | 231.90(3) |
| x =0.40 | 5.482(4) | 5.471(4) | 7.736(4) | 2.19 | 232.08(3) |
| x =0.50 | 5.474(4) | 5.482(4) | 7.738(4) | 2.07 | 232.29(3) |
| x =0.70 | 5.490(4) | 5.478(4) | 7.745(4) | 2.11 | 232.98(3) |
| x =1.00 | 5.503(4) | 5.499(4) | 7.771(4) | 2.26 | 235.20(3) |

**Table 2**

| Composition $L_{0.7}Ca_{0.3}Mn_{1-x}Fe_x O_3$ | Activation energy (meV) | $T_{MI}$ (K) | $\theta_D$ (K) | $T_C$ (K) | | $M(\mu_B)$ Experimental | $M(\mu_B)$ Calculated |
|---|---|---|---|---|---|---|---|
| x =0.00 | 49.2 | 257 | 520 | 250.30 | | 3.96 | 3.70 |
| x =0.02 | 51.8 | 229 | 466 | 215.30 | | 3.97 | 3.57 |
| x =0.04 | 50.7 | 193 | 396 | 179.85 | | 3.89 | 3.45 |
| x =0.06 | 47.4 | 151 | 322 | 139.72 | | 3.625 | 3.32 |
| x =0.08 | 37.4 | 90 | 220 | 124.49 | | 3.072 | 3.20 |
| x=0.10 | - | - | - | 104.80 | | 2.553 | 3.07 |
| x=0.20 | - | - | - | 60.95 | | 1.032 | 2.44 |
| x=0.30 | - | - | - | 50.74 | | 0.513 | 1.81 |

*******************************